# Multi-window SRS imaging using a rapid widely tunable fiber laser

Hongli Ni,[1,‡] Peng Lin,[1,‡] Yifan Zhu,[2] Meng Zhang,[3] Yuying Tan,[3] Yuewei Zhan,[3] and Ji-Xin Cheng[1,2,3,*]

[1]Department of Electrical and Computer Engineering, Boston University, 8 St. Mary's St., Boston, MA, 02215, USA
[2]Department of Chemistry, Boston University, 590 Commonwealth Avenue, Boston, MA 02215, USA
[3]Department of Biomedical Engineering, Boston University, 44 Cummington Mall, MA 02215, USA

**ABSTRACT:** Spectroscopic stimulated Raman scattering (SRS) imaging has become a useful tool finding a broad range of applications. Yet, wider adoption is hindered by the bulky and environmentally-sensitive solid-state optical parametric oscillator (OPO) in current SRS microscope. Moreover, chemically-informative multi-window SRS imaging across C-H, C-D and fingerprint Raman regions is challenging due to the slow wavelength tuning speed of the solid-state OPO. In this work, we present a multi-window SRS imaging system based on a compact and robust fiber laser with rapid and widely tuning capability. To address the relative intensity noise intrinsic to fiber laser, we implemented auto-balanced detection which enhances the signal-to-noise ratio of stimulated Raman loss imaging by 23 times. We demonstrate high-quality SRS metabolic imaging of fungi, cancer cells, and *Caenorhabditis elegans* across the C-H, C-D and fingerprint Raman windows. Our results showcase the potential of the compact multi-window SRS system for a broad range of applications.

Coherent Raman Scattering (CRS) microscopy enables high-speed chemical mapping through probing intrinsic molecular vibrations. CRS spectroscopic imaging overcomes the speed limitation of spontaneous Raman spectroscopy and has significantly extended the scope of applications in the past two decades [1, 2]. A CRS microscope utilizes two ultrafast lasers, named pump at $\omega_p$ and Stokes at $\omega_S$, to coherently excite molecular vibrations at the two lasers' beating frequency ($\omega_p-\omega_S$). CRS microcopy can be realized by coherent anti-Stokes Raman scattering (CARS) at the wavelength of $2\omega_p-\omega_S$, or stimulated Raman scattering (SRS) manifested as the intensity loss at pump beam (stimulated Raman loss, SRL) and the intensity gain of Stokes (Stimulated Raman gain, SRG). SRS has the advantages over CARS in that it offers nearly identical spectral profiles as spontaneous Raman spectroscopy. SRS signal has a linear relationship to the molecule concentration, thus allowing quantitative measurement [3]. Furthermore, SRS is measured under ambient light, which is preferable for clinical use.

The above-mentioned advantages have made SRS microscopy an increasingly useful tool in various biomedical and clinical research [4-9]. However, broader adoption of this technique is challenged by the routinely-used solid-state optical parametric oscillator (OPO) laser due to its bulkiness and strict requirement for the operating environment. Furthermore, the slow wavelength tuning speed of solid-state OPO hinders multi-window SRS imaging, i.e., spectroscopic SRS imaging across multiple Raman windows including C-H, C-D and fingerprint region.

Current spectroscopic SRS imaging methods can be categorized into two groups according to the bandwidth of the laser pulse. The first category uses narrowband picosecond pulses and generates a spectrum by tuning the laser wavelength, where high-throughput imaging is limited by the slow wavelength tuning of the solid-state OPO [8-10]. The second category uses broadband femtosecond laser and harnesses spectral focusing [11-13] or other pulse shaping [14-16] techniques for hyperspectral imaging within the bandwidth of the laser pulses. Such approaches enable high speed spectroscopic imaging within a limited spectral window, typically less than 300 cm$^{-1}$ when using 100 fs pulses. Multi-window imaging has to rely on tuning the solid-state OPO, which is slow and suffers from laser pointing drift.

To overcome the limitations of the solid-state OPO, fiber lasers have gained increasing interest in SRS microscopy for its compactness and robustness. Many dual-output fiber lasers have been developed and deployed for SRS imaging, including synchronized fiber laser and Ti:sapphire laser [17, 18], electronically synchronized fiber lasers [19], self-synchronized fiber lasers[20], supercontinuum generation [21, 22], soliton frequency self-shifting [23], Fourier domain mode-locking [24], and parametric frequency generation [25]. Nevertheless, the above-mentioned fiber laser sources either have limited spectral coverage or low power spectral density, which are not suitable for multi-window spectroscopic SRS imaging.

Recently, Brinkmann *et al* developed a rapid widely tunable picosecond fiber laser based on a fiber-OPO (FOPO) and deployed this laser for CARS imaging [26, 27]. By engineering the dispersion and nonlinearity of the fiber, the FOPO enables broad spectral tuning which covers C-H, C-D and fingerprint region. Through incorporating a chirped fiber Bragg grating to match the repetition rate of the two outputs automatically, this FOPO achieves a spectral tuning speed of 5 ms between arbitrary wavenumbers, which is beyond the reach by other fiber lasers used for coherent Raman microscopy.

In this paper, we demonstrate SRG and SRL imaging with the rapid widely tunable FOPO. One challenge for fiber laser-based SRS is the intrinsic high laser relative intensity noise (RIN). We therefore analyzed the noise spectrum of the fiber laser and optimized the modulation frequency for SRG imaging. Next, we implemented auto-balanced detection (ABD) on the FOPO output, which improves the signal-to-noise ratio (SNR) by 23 times for SRL imaging. Subsequently, we demonstrated SRL imaging of biological specimens cross the C-H, C-D and fingerprint regions. In the C-H region, we mapped the lipid and protein distribution inside fungi and ovarian cancer cells. Through cross-window analysis of the C-D and C-H region, we visualized the metabolic activity of fungi and lipid uptake of ovarian cancer cells. Combining the fingerprint and C-H region images, we located the unsaturated lipid-rich as well as cholesterol-rich organs and generated the lipid unsaturation ratio map of the *Caenorhabditis elegans (C.elegans)*. Collectively, the results demonstrate the potential of our multi-window SRS microscope for a wide range of applications.

## METHODS

**Multi-window SRS imaging system.** Fig. 1 (a) shows the schematic. The fiber laser (PICUS DUO, Refined Laser Systems GmbH) generates two synchronized pulse trains through the FOPO and an ytterbium-doped fiber oscillator. The pulses from the FOPO serves as the pump and the pulses from the fiber oscillator serves as the Stokes for SRS imaging. The laser repetition rate is 40.5 MHz and the pulse duration is ~7 ps. The spectral tuning range is from 1050 $cm^{-1}$ to 3150 $cm^{-1}$, corresponding to pump wavelength from ~946 nm to 775 nm and Stokes wavelength from ~1053 nm to 1027 nm (Supplementary Fig. S1). The spectral resolution is about 12 $cm^{-1}$. The timing of the two beams varies for different Raman windows. To ensure the temporal overlapping of the pump and Stokes pulses, a motorized stage (X-LSM025A-KX14A, Zaber technology) is used to compensate the timing change for every 300 $cm^{-1}$.

The pump and Stokes beams first pass a half-wave plate (HWP) and a polarization beam splitter (PBS) for power adjustment and polarization parallelization. In the pump path, a 750 nm longpass filter is put after the PBS to filter out possible stray light. Both beams are pre-expanded to around 2 mm in diameter by a 4-f system. Between the 4f lens pair, an acousto-optic modulator (AOM) (1205-C, Isomet) is set near the beam focus to modulate the laser beam. For SRG, the pump is modulated and for SRL the Stokes. The pump and Stokes beams are spatially combined on the dichroic mirror. The combined beams then pass a pair of galvo mirrors for laser X-Y scanning. A 4-f system is between the galvo mirror and the objective to expand the laser beam to around 6 mm in diameter and conjugate the galvo mirror with the objective back aperture. The spatio-temporally overlapped pump and Stokes beams are focused onto the sample by a water-immersion objective with 1.2 numerical aperture (NA) (UPLSAPO60XW, Olympus). A piezo positioner is installed on the objective to enable axial scanning. The outgoing light is collected by an oil immersion condenser with 1.4 NA (U-TLO, Olympus). The collected light passes a spectral filter to filter out the AOM-modulated beam. For SRL we used a 1000 nm shortpass filter (FESH1000, Thorlabs) and for SRG we used a 1000 nm longpass filter (FESH1000, Thorlabs).

The filtered light is detected by a home-built detector (Fig. 1 (b)). The light signal is transformed to photocurrent by a large area 62 V-biased photodiode (DET100A2, Thorlabs). The photocurrent is then turned to voltage signal after passing a 50 Ω impedance. The AC and DC components of the signal is separated by a bias tee (ZFBT-282-1.5A+, Mini-circuits). The separated DC is sent out as the DC output. The AC part is pre-filtered by 1.2 MHz highpass (ZFHP-1R2-S+, Mini-circuits) and 22 MHz lowpass filter (BLP-21.4+, Mini-circuits) then amplified by a low-noise voltage amplifier (SA-230F5, Wayne Kerr Electronics). The amplified AC signal is the AC output of the detector. When imaging without the ABD, this AC output is directly demodulated by a lock-in amplifier (HF2LI, Zurich Instrument) to extract the SRS signal.

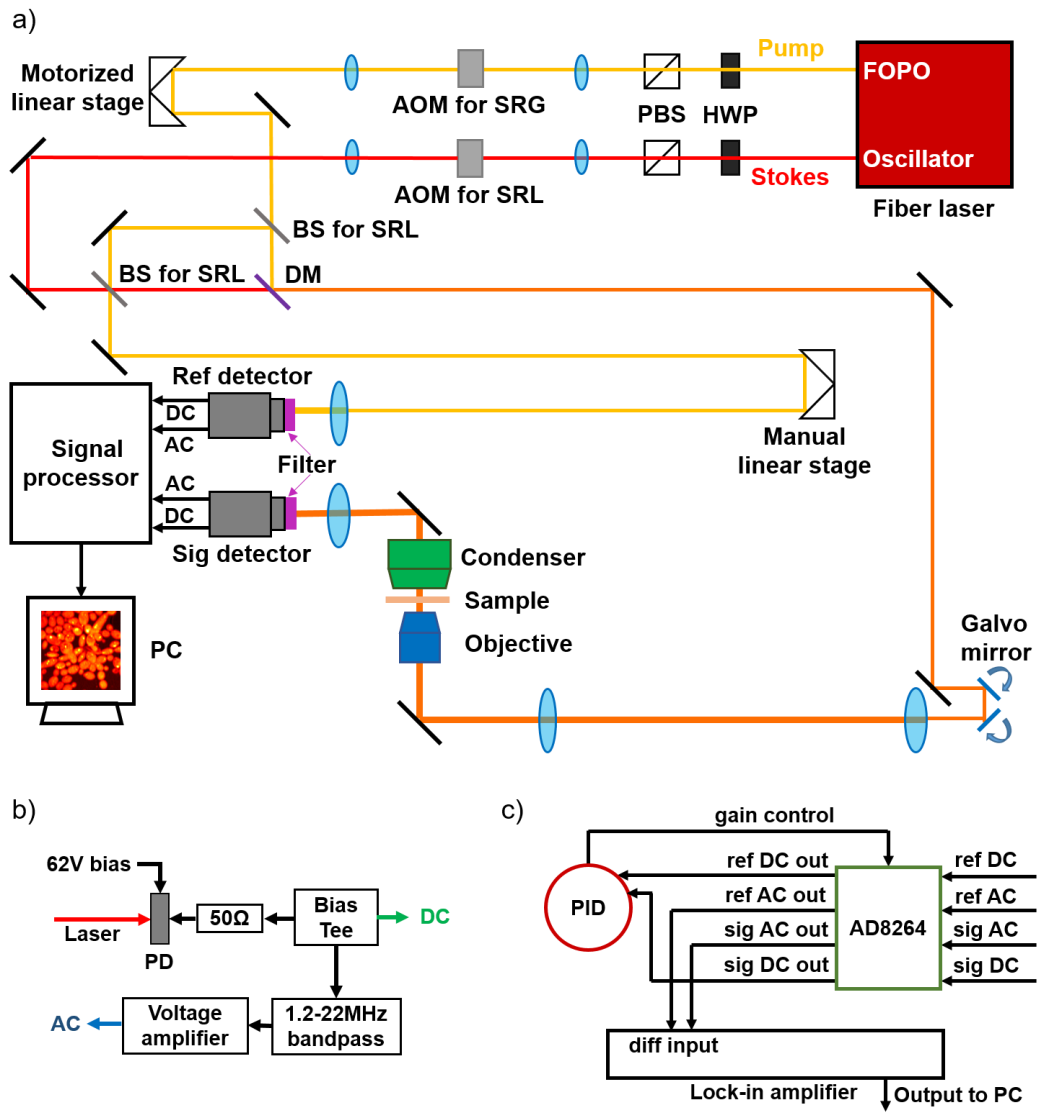

**Fig. 1.** A rapid, widely tuning SRS microscope. (a) Schematic. HWP: Half-wave plate. PBS: Polarization beam splitter. AOM: Acousto-optic modulator. BS: Beam splitter. DM: Dichroic mirror. Ref: Reference. Sig: Signal. (b) Schematic of the detector assembly. PD: Photodiode c) Workflow inside the signal processor for auto-balanced detection. PID: proportional–integral–derivative controller.

**Auto-balanced detection**. In balanced detection, the noisy laser beam is split into a signal and a reference arm. The signal arm interacts with the sample and the reference arm does not. The laser noise is canceled by taking the difference of the electrical signal from the two arms. The electrical signal from the two arms is carefully balanced in phase and amplitude to obtain the best noise cancellation performance. Compared to the optical delay-based [19] or polarization-based [28, 29] method, ABD requires no special optics as well as fine optical alignment. ABD utilizes a proportional–integral–derivative (PID) controller to balance the signal amplitude of the two arms, making it robust for sample transmission variation [15, 22].

To implement auto-balanced SRS, several components are added into the system (Fig. 1 (a)). A beam splitter (BS) (BSS11R, Thorlabs) is put into the light path to split the unmodulated beam into 2 arms. The two arms are detected by identical detectors as illustrated in Fig. 1(b). The detector for the signal arm is denoted as sig detector and the one for the reference arm is denoted as ref detector. The phase of the AC signal from the two arms is matched by adjusting the manual linear stage in the reference arm (Supplementary Fig. S2). The AC amplitude is balanced with the auto-balancing signal processor shown in Fig. 1 (c). In the signal processor, the AC and DC outputs from the two detectors are sent to a 4-channel variable gain amplifier (AD8264, Analog Devices Inc.). The function of AD8264 is to amplify the signals with adjustable gain. To balance the AC amplitude, the gain for the two arms is controlled by a PID controller (Moku:Lab, Liquid Instruments). The PID determines the gain according to the DC signal from the two arms.

**Samples.** The fungal strain, C. albicans SC5314 (wild type), used in this study was from American Type Culture Collection (ATCC). *C. albicans* cells were grown overnight in yeast extract peptone dextrose (YPD) at 30 °C in an incubating shaker to reach stationary phase. The stationary-phase fungal cells were then diluted by 20 times in fresh YPD and cultured for another 5 hours at 30°C to obtain the log-phase cells. After that, fungal specimens were pelleted and washed with 1×PBS. The log-phase *C. albicans* cells were resuspended in 0% and 90% D₂O containing YPD for D₂O treated group and control group at a concentration of 1×10⁷ cells/mL and incubated for 2 hours. Then, 1 mL of fungi solution was centrifuged, washed with 1×PBS, and fixed by 10% formalin solution. Before imaging, formalin solution was washed away by 1×PBS.

The ovarian cancer cells OVCAR5 were provided by Daniela Matei lab at Northwestern University and were cultured in RPMI 1640 medium (Gibco) supplemented with 10% FBS, 1% P/S and 2 mM GlutaMax (Gibco). Cells were seeded in 35 mm glass-bottom dishes (Cellvis) overnight and were fixed with 10% neutral buffered formalin for 30 min followed by 3 times PBS wash before imaging. For oleic acid-d34 labeling, cells were cultured with 100μM oleic acid-d34 (Cambridge Isotope Lab) for 6 hours after overnight seeding in the glass-bottom dishes. All cells were maintained at 37 °C in a humidified incubator with 5% CO2 supply.

The *C. elegans* daf-2 (e1370) mutant strain was used in SRL imaging. The *C. elegans* was grown on standard NGM plates (containing 5 mg/L cholesterol) with *E. coli* OP50 at 20 °C for 2-3 days. Both daf-2 mutant *C. elegans* and *E. coli* OP50 were purchased from the Caenorhabditis Genetics Center at University of Minnesota. Before imaging, the *C.elegans* was detached from NGM plate and fixed in 10% formalin.

**Spontaneous Raman spectroscopy.** The Raman spectra were measured using a commercial Raman spectrometer (LabRAM HR, Horiba).

## RESULTS

**Laser noise measurement and SRG imaging performance.** We measured the RIN spectra of the fiber oscillator and the FOPO outputs using the setup sketched in Fig. 2 (a). The detector used for RIN measurement is the same as the one described in section 2.1. The AC output is sent to a spectrum analyzer (USB-SA44B, Signal Hound) to obtain the noise spectrum $S(f)$. The unit for $S(f)$ is dBm, which is a power unit. The value of $S(f)$ is influenced by the noise equivalent bandwidth (NBW), the signal amplification gain $G_{AC}$ and the laser power on detector. To better compare the noise level, the noise is normalized to get the RIN spectrum [30]. We used the equation (1) to calculate RIN spectrum. The $V_{DC}$ is the voltage of the detector DC output and is measured by a voltmeter. The $\frac{V_{DC}^2}{R}$ is the DC power, which represents the influence of average laser power on detector.

$$RIN = \frac{S(f)}{NBW * G_{AC} * \frac{V_{DC}^2}{R}} \quad (1)$$

Fig. 2 (b) gives the laser RIN result with unit dBc/Hz. A RIN spectrum measured from a solid-state OPO (InsightX3, Spectra-Physics) is plotted as a reference. The RIN result indicates that the FOPO output is much noisier than the oscillator output. Therefore, we first performed SRG imaging to avoid the large laser noise in the FOPO output. We also noticed that the oscillator output has lower noise in higher frequency, suggesting that the modulation frequency (MF) can be increased to reduce the noise level in SRG imaging.

Fig. 2 (c) shows the SRG images of 3 μm PMMA beads. We defined the SNR of a SRS image by equation (2), where the signal mean is measured from the area with strong SRS signal and the noise standard deviation (std) is measured from the background.

$$SNR = \frac{Signal\ mean}{noise\ std} \quad (2)$$

The SNR of an image with 6MHz MF is 2 times of that with 1.5 MHz MF, matching with the RIN spectrum in Fig. 2 (b). Due to the limited AOM bandwidth, the modulation waveform deteriorates as the MF increases, resulting in SRS signal level drop in the higher MF (Supplementary Fig. S3 (a)). The sweet point for the SNR is around 6MHz (Supplementary Fig. S3 (b)).

Yet, the optimized SRG modality is not enough for cellular imaging at the C-D and fingerprint regions. Fig. 2(d) shows the SRG images of D₂O treated fungi cells with 6 MHz MF at 2935 cm⁻¹ (C-H region) and 2150 cm⁻¹ (C-D region). The C-H region image barely shows the cell morphology. At the C-D region, the SNR is too low to show any contrast.

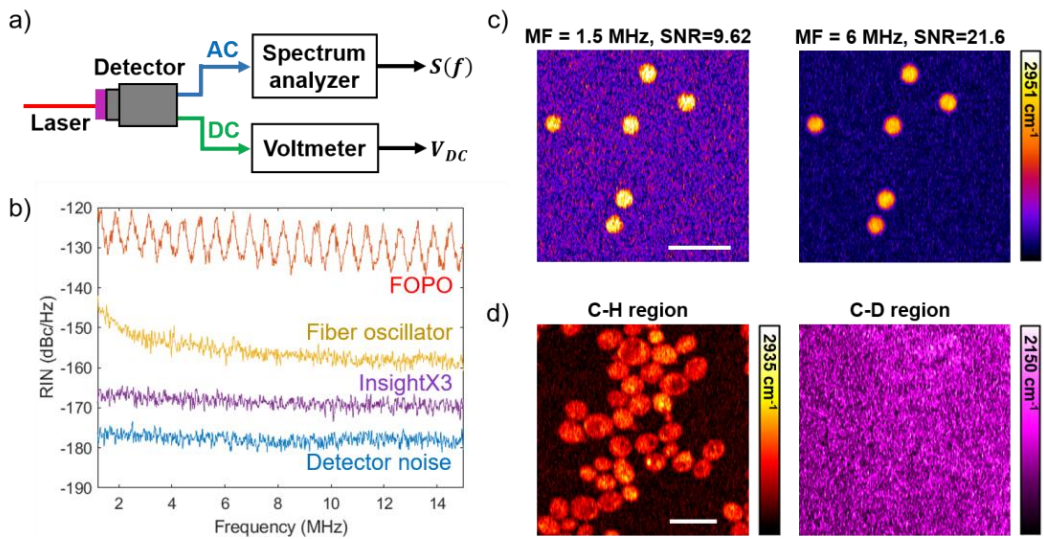

**Fig. 2.** Laser noise measurement and SRG imaging result. (a) Laser RIN measurement setup. (b) Laser RIN measurement result. Laser power on detector: ~20mW. (c) SRG imaging of 3 μm PMMA beads. MF: modulation frequency. Power on sample: Pump: 24 mW, Stokes: 22 mW. Dwell time: 10 μs. (d) SRG imaging of $D_2O$ treated fungi at 2935 $cm^{-1}$ (C-H region) and 2150 $cm^{-1}$ (C-D region). Power on sample: C-H: Pump: 18 mW, Stokes: 43 mW. C-D: Pump: 30 mW, Stokes: 40 mW. Dwell time: 100 μs. MF: 6MHz. Scale bar: 10 μm.

**Auto-balanced SRS imaging performance.** To further improve the SNR, we implemented the ABD to cancel the laser intensity noise. Fig. 3 (a) shows the oscillator noise spectrum before and after ABD, measured by the lock-in oscilloscope function. Fig. 3 (c) shows the FOPO noise spectrum before and after ABD. With 10 mW laser power on the detector, ABD suppressed the oscillator noise by ~4 dB and the FOPO noise by ~26 dB at 3.3 MHz (Fig. 3 (a) and (c)). This noise reduction led to SRG SNR improvement by 1.59 times and SRL SNR improvement by 20 times. Fig. 3 (b) and (d) compares the SRS imaging result of 3 μm PMMA beads with and without ABD. The modulation frequency of 3.3 MHz was chosen to avoid too much signal level drop caused by the modulation deterioration (Supplementary Fig. S3). The SNR of the SRG image is improved by ~1.58 times and that of the SRL image is improved by ~23 times. The SNR improvement is more significant for SRL as the FOPO output is much noisier than the oscillator output. The auto-balanced SRL achieves better SNR than the auto-balanced SRG, mainly because the oscillator output has higher power than the FOPO output.

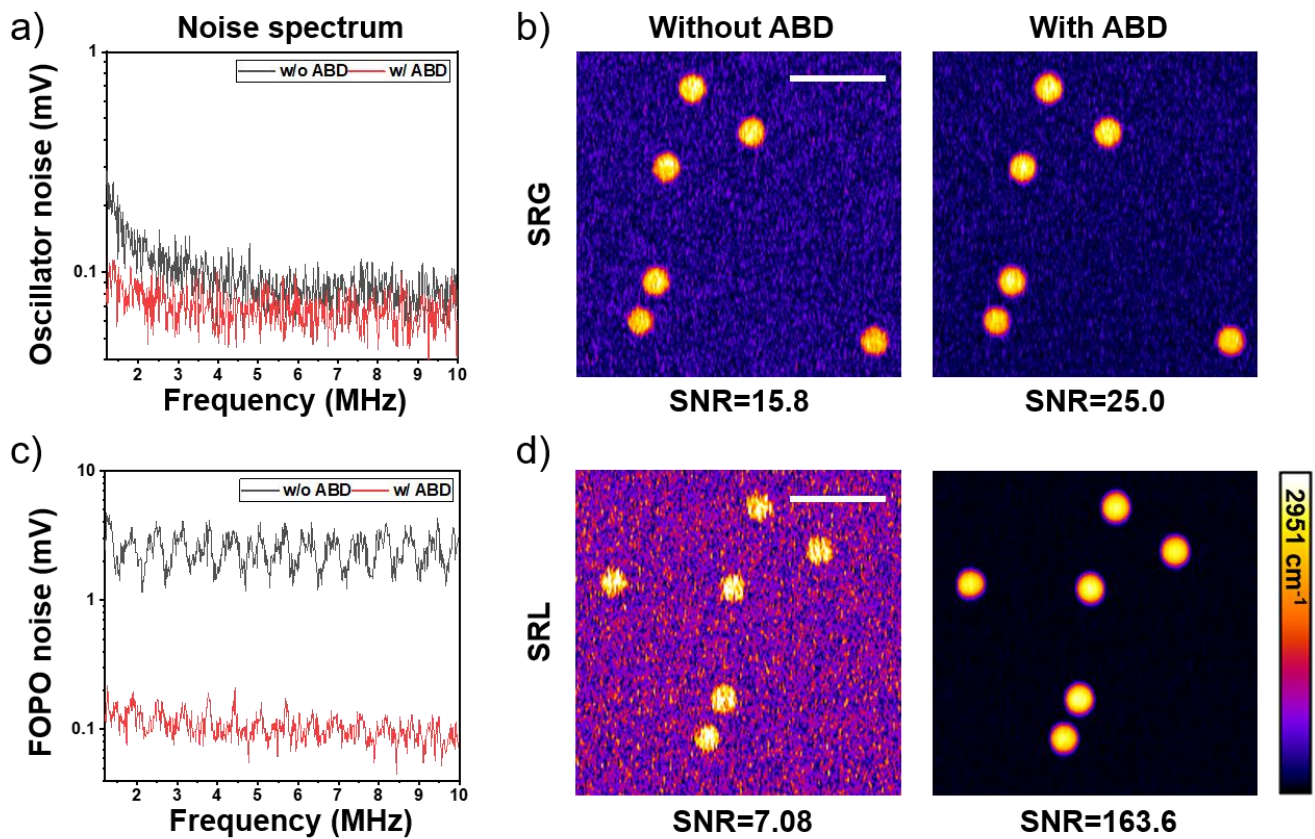

**Fig. 3.** Auto-balanced detection (ABD) for improving SNR of SRS imaging. (a) Oscillator noise spectrum with and without ABD. Power on PD: 11 mW. (b) SRG imaging of 3 μm PMMA beads at 2951 cm$^{-1}$ captured with and without ABD. Power on sample: Pump: 18 mW, Stokes: 22 mW. (c) FOPO noise spectrum with and without ABD. Power on PD: 10 mW. (d) SRL imaging of 3 μm PMMA beads at 2951 cm$^{-1}$ captured with and without ABD. Power on sample: Pump: 16 mW, Stokes: 99 mW. Dwell time: 10 μs. Scale bar: 10 μm.

With the auto-balanced SRL, we characterized the spectral fidelity, detection sensitivity, and excitation volume. Fig. 4 (a) shows SRL spectra of several standard samples in C-H, C-D and fingerprint regions. The overall shape of the SRL spectra matches well the spontaneous Raman data (Fig. 4 (a) dash line). The spectra in Fig. 4 (a) were post-processed to remove artificial sharp dips caused by the laser imperfections (Supplementary Fig. S4 (a)). The post-processing procedure manually deletes the bad data points and then smooths the spectra using moving average with a window of 5 data points (Supplementary Fig. S4 (c)). Because the raw SRL spectra were sampled with 1 cm$^{-1}$ step, which is much smaller than the spectral resolution, the resolution of the SRL spectrum should not be affected by the post-processing. Fig. 4 (b) presents the detection sensitivity of DMSO at 2912 cm$^{-1}$. The detection sensitivity is found to be better than 0.02 mol/L. Fig. 4 (c) and (d) depicts the transverse and longitudinal view of 3D SRL imaging of a 500 nm PMMA bead. The intensity profiles along the lateral and axial direction are shown in Fig. 4 (e) and (f). The system axial resolution is around 2.47 μm and the lateral resolution is estimated to be 392 nm after deconvolution with particle size.

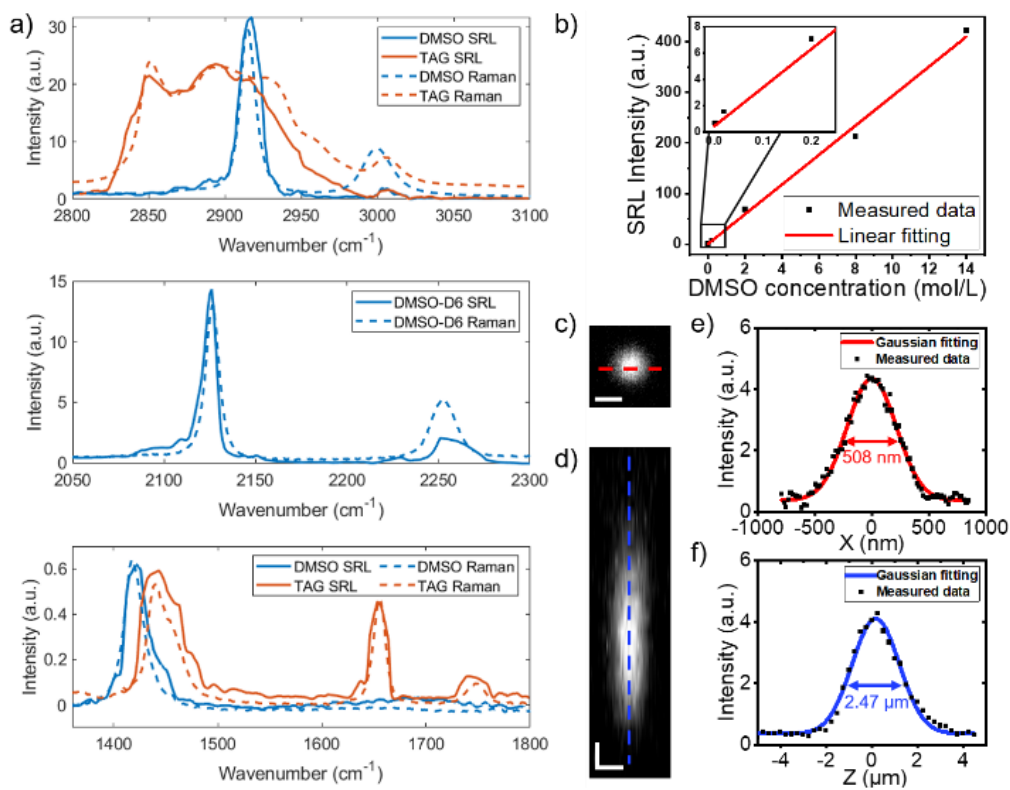

**Fig. 4.** Characteristics of auto-balanced SRL. (a) SRL spectra of representative samples in C-H, C-D, and fingerprint vibration regions. (b) DMSO detection sensitivity. (c) Transverse view of 3D SRL imaging of a 500-nm PMMA bead. (d) Longitudinal view of 3D SRL imaging of a 500-nm PMMA bead. (e) Intensity profile along the red line in (c). (f) Intensity profile along the blue line in (d). Power on sample: pump: 15 mW, Stokes: 90 mW. Dwell time: 100μs. Scale bar: 500 nm.

**Multi-window SRL imaging of biological samples.** To evaluate the imaging performance in the C-H region, we first recorded the SRL spectra of bovine serum albumin (BSA) and triacylglyceride (TAG), as shown in Fig. 5 (a). We applied the same post-processing method to remove the artificial sharp dips in the SRL spectra as mentioned above. Next, we took hyperspectral image stacks of fungi and ovarian cancer cells from 2800 to 3100 cm$^{-1}$ with 1 cm$^{-1}$ step size. The spectra at every pixel are processed in the same way as the BSA and TAG spectra. The average results of the hyperspectral stacks are shown in the spectral projection column in Fig. 5 (b) and (c). By applying a least absolute shrinkage and selection operator (LASSO) spectral unmixing algorithm [11] and taking the BSA and TAG spectra as protein and lipid spectra reference, maps of protein and lipid distribution in the cells are obtained from the hyperspectral data, as shown in Fig 5. (c) and (d). The lipids are located in the cytoplasm, while the protein signals were found in both nucleus and cytoplasm.

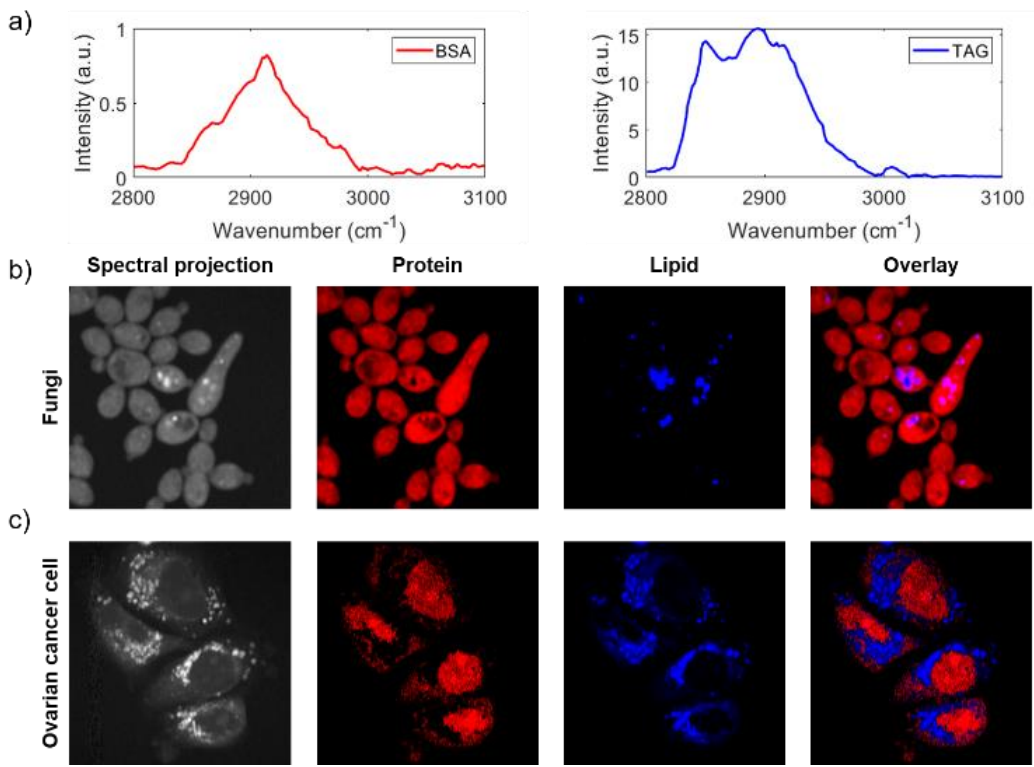

**Fig. 5.** SRL imaging of cells in the C-H region. (a) Standard sample spectra used for LASSO. (b) Hyperspectral stack projection and spectral unmixing result of fungi cells. (c) Hyperspectral stack projection and spectral unmixing result of ovarian cancer cells. Power on sample: Pump: 16 mW, Stokes: 98 mW. Dwell time: 10 μs. Scale bar 10 μm

Tracking the uptake of deuterium-labeled nutrients using SRS imaging is an important tool for cell metabolism study [6, 8]. When deuterium-labeled nutrients are incorporated into the biomass, part of the C-H bonds in the cell molecules is replaced by the C-D bond, which gives signature Raman peaks in the C-D region. The SRS signal ratio of $I_{C-D}/(I_{C-D}+I_{C-H})$ therefore reflects the activeness of cell metabolism [31]. Due to the slow speed in switching between C-D region and C-H region, most of the published SRS studies only used the C-D intensity to represent cell activity [6, 8]. Yet, the C-D SRS intensity can be influenced by laser intensity, focus drifting, system alignment etc., thus not robust enough to quantify the cell metabolism. Here, multi-window SRS imaging addresses this issue with the capability of rapidly switching between the C-D and C-H vibrations. C-D/C-H SRL imaging of deuterium-labeled fungi and ovarian cancer cells are shown in Fig. 6 a-d. The $D_2O$ treated fungi cells can be clearly seen in the C-D channel. In the contrast, the control group only shows very vague cell shapes due to cross phase modulation (Fig. 6 (a)). Similarly, the perdeuterated oleic acid (OA-d34) treated ovarian cancer cells have strong C-D signal from the lipid, while the control group has no signal (Fig. 6 (d)). The C-H SRS images of the same cells are recorded as references. The heat map of $I_{C-D}/(I_{C-D}+I_{C-H})$ ratio for the fungi and ovarian cancer cells are shown in Fig. 6 (c) and (f), where the metabolism heterogeneity of the microbes and the lipid accumulation in ovarian cancer cells are clearly seen.

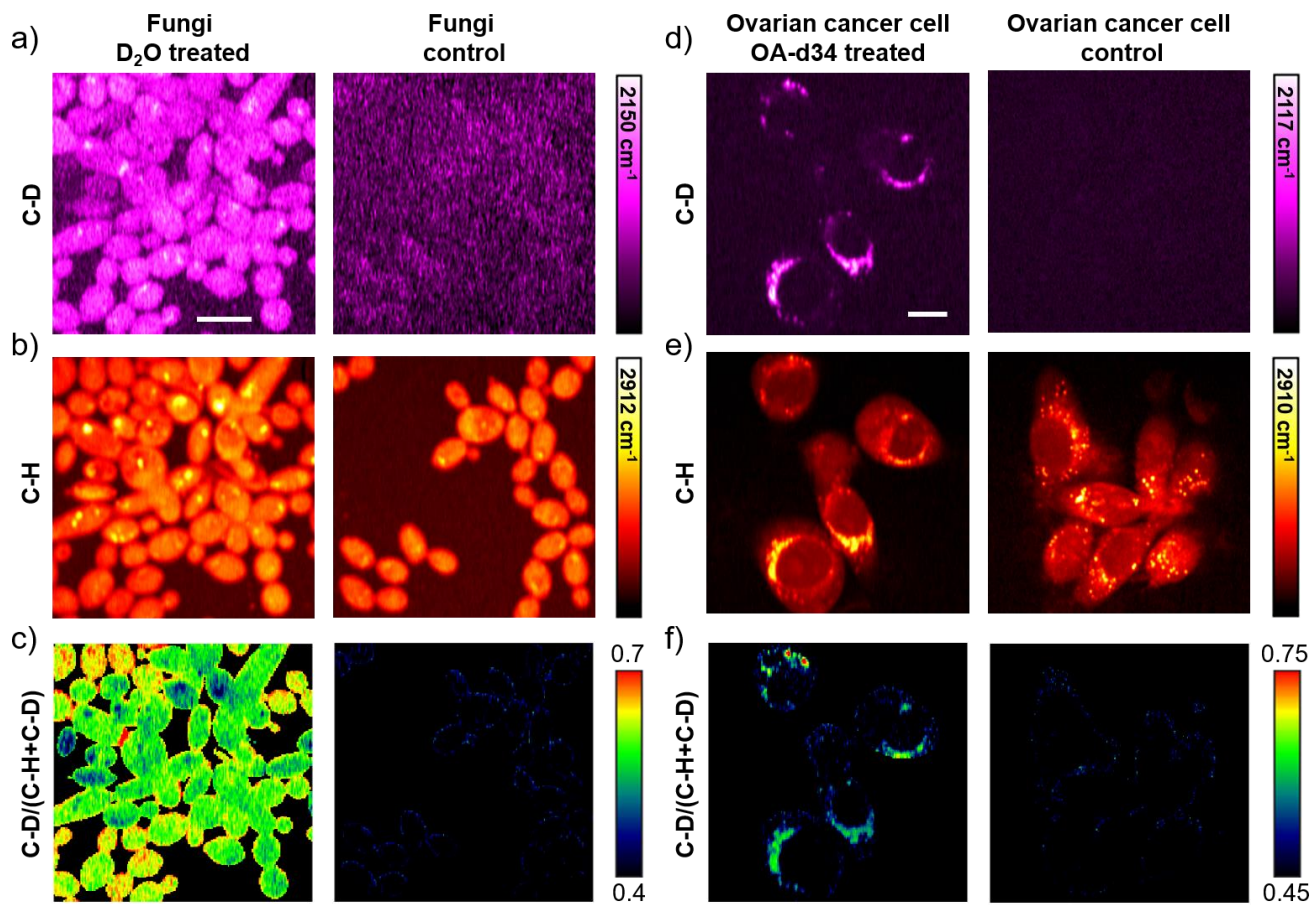

**Fig. 6.** C-D/C-H SRL imaging of fungal and cancer cells. (a) C-D SRL image of fungi cells with and without $D_2O$ treatment. Power on sample: Pump: 21 mW, Stokes: 93 mW. Dwell time: 100 μs. 40 average. (b) C-H SRL image of fungi cells with and without $D_2O$ treatment. Power on sample: Pump: 16 mW, Stokes: 95 mW. Dwell time: 10 μs. 10 average. (c) Heat map of $I_{C-D}/(I_{C-D}+I_{C-H})$ ratio calculated using (a) and (b). (d) C-D SRL image of ovarian cancer with and without OA-d34 treatment. Power on sample: Pump: 21 mW, Stokes: 93 mW. Dwell time: 100 μs. 10 average. (e) C-H SRL image of ovarian cancer with and without OA-d34 treatment. Power on sample: Pump: 16 mW, Stokes: 95 mW. Dwell time: 10 μs. 10 average. (f) Heat map of $I_{C-D}/(I_{C-D}+I_{C-H})$ ratio calculated using (d) and (e). Scale bar 10 μm.

Lastly, to evaluate the performance of our microscope for organism study, we imaged the same *C.elegans* in both fingerprint and C-H region. We chose 3 representative spectral channels: the 1650 cm$^{-1}$ channel targets the acyl C=C bond and shows the distribution of unsaturated fatty acids; the 1670 cm$^{-1}$ channel corresponds to the sterol C=C bond and reflects the cholesterol distribution; The 2845 cm$^{-1}$ in the C-H region shows the overall lipid storage. By thresholding the SRL images at 1650 and 1670 cm$^{-1}$, we located the unsaturated lipid-rich organs and cholesterol-rich organs (Fig. 7 (b)). We also calculated the lipid unsaturation ratio of the fat droplets using the equation $I_{acyl\ C=C}/(I_{acyl\ C=C}+I_{C-H})$. The heat map of the lipid unsaturation is shown in Fig. 7 (c), which depicts heterogeneity of the fat droplets inside the worm. We also acquired depth-resolved images of acyl and sterol C=C bonds (Fig. 7(b)) and the lipid unsaturation map (Fig. 7(c)). The depth-resolved images allow quantitative analysis of lipid metabolism.

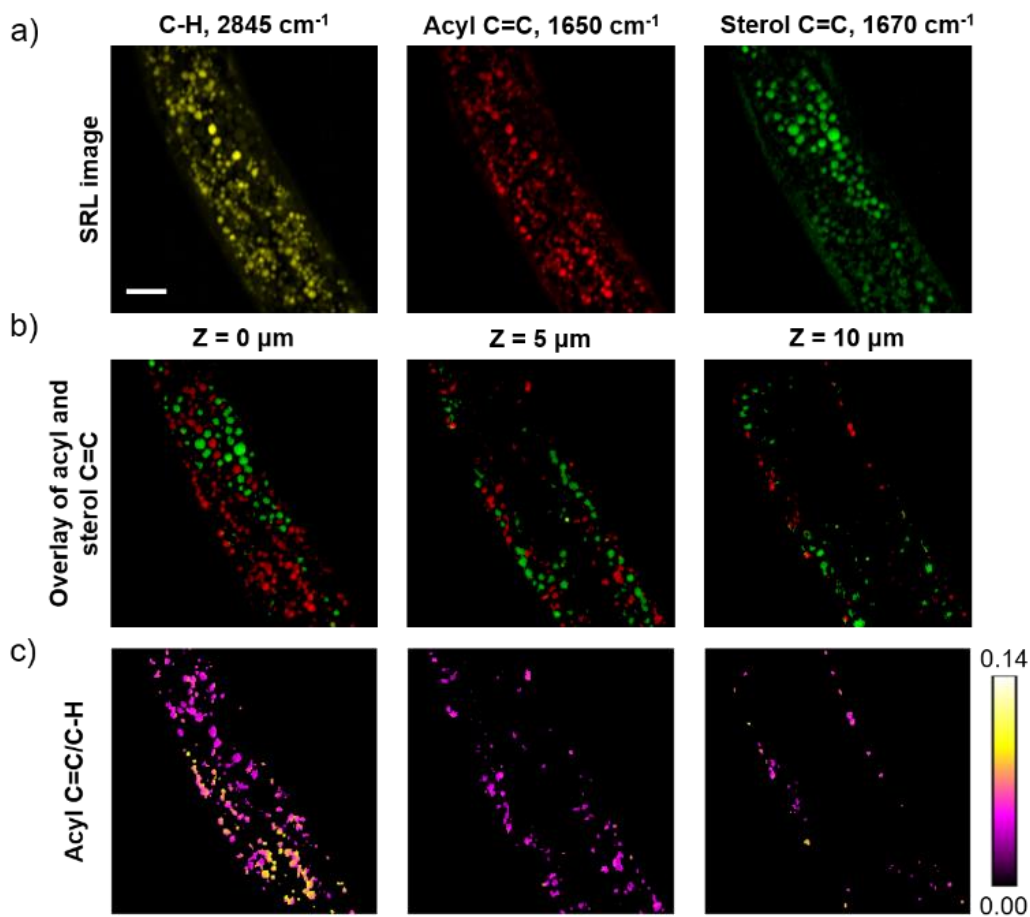

**Fig. 7.** C-H and fingerprint SRL imaging of *C. elegans*. (a) SRL images at 2845, 1650 and 1670 cm$^{-1}$. Power on sample: 2845 cm$^{-1}$: Pump: 16 mW, Stokes: 99 mW. 1650 cm$^{-1}$ and 1670 cm$^{-1}$: Pump: 21 mW, Stokes: 88mW. Averaging: 2845 cm$^{-1}$: no average. 1650 cm$^{-1}$: 10 average. 1670 cm$^{-1}$: 20 average. Dwell time: 100 µs. Scale bar 10 µm. (b) Unsaturated lipid-rich and cholesterol-rich organs distribution. This distribution is generated by thresholding SRL images at 1650 cm$^{-1}$ and 1670 cm$^{-1}$ and then merging the two channels. (c) Heat map of lipid unsaturation ratio for lipid droplets.

## DISCUSSION

We demonstrated multi-window SRS imaging with a rapid widely tunable fiber laser. The fast spectral tuning capability across C-H, C-D and fingerprint regions enables cross-window quantitative analysis of cell metabolic activity. Furthermore, our multi-window SRS microscope is able to sample the spectrum selectively and discretely, which is in the contrary to spontaneous Raman [32] and broadband CARS imaging [33]. As the spectral data is highly sparse [34], the selective sampling would greatly boost the spectroscopic imaging speed and meanwhile lower the total laser energy on the sample.

In this work, we compared the SRG and SRL modalities. We note an independent, parallel work by Würthwein et al. demonstrated SRG imaging of polymer beads (30 µm in diameter), pure dDMSO (14.1 mol/L), lipid and dDMSO-soaked adipose tissue using the same laser source [35]. Due to limited power of the FOPO output (pump beam in SRS), SRG imaging of cell metabolism remains difficult (see Fig. 2 (d)). By applying auto-balanced detection to the FOPO beam, we successfully developed an SRL modality with greatly enhanced detection sensitivity (i.e., 20 mM DMSO as shown in Fig. 4 (b)). With this modality, we have mapped chemicals in biological cells and *C. elegans* in C-H, C-D, and fingerprint vibration windows.

In the fingerprint and C-D region, we applied a relatively long dwell time and multiple averaging to get good SNR, resulting in an imaging speed of 1~2 minutes per frame. The rapid spectral tuning speed of the laser is not fully utilized at this imaging speed. The SNR can be further improved with several approaches. Firstly, it is possible to suppress the laser noise to a lower level with the ABD. By modeling the residual noise after ABD suppression, we found the SRS detection is not shot-noise-limited or electrical noise-limited (Supplementary Fig. S5). In parallel, it is possible to further optimize the ABD module to achieve shot-noise limit detection. The second approach is to increase the detector saturation power threshold. The detector saturation issue limits the maximum laser power for the unmodulated beam. We will test photodiodes that have higher saturation level. By using new detectors with higher saturation level, we can potentially use full laser power provided by the FOPO and the oscillator, which could improve the SRL SNR by ~2 times and SRG SNR by ~9 times.

The FOPO instability is another issue of the current multi-window SRS system, which resulted in artificial sharp dips in current SRL spectra. This issue can be solved within a small tuning range by not tuning the oscillator wavelength but the repetition rate. With a fixed oscillator wavelength, the system allows tuning the FOPO wavelength over a small range in a smooth way, as demonstrated by Würthwein et al. using the same laser source [35].

## CONCLUSIONS

We reported a multi-window SRS imaging system enabled by a rapid and widely tunable fiber OPO and auto-balanced detection. Our system is able to access Raman bands from 1050 cm$^{-1}$ to 3150 cm$^{-1}$ and can tune between two arbitrary wavenumbers within 5 ms. By analyzing the laser noise spectrum, we optimized the modulation frequency for SRG imaging. Further, we demonstrated auto-balanced SRL imaging with much improved SNR. By implementing auto-balanced detection, we suppressed the pump noise by ~26 dB at 3.3 MHz, leading to an SNR improvement of 23 times. With the auto-balanced SRL, we demonstrated multi-window SRS imaging across the C-H, C-D and fingerprint regions. Through hyperspectral SRL imaging in C-H region, lipid and protein distribution inside fungi and ovarian cancer cells are mapped. With C-D/C-H SRL imaging, $D_2O$ metabolism in fungi and fatty acid uptake by ovarian cancer cells are visualized. Combining fingerprint and C-H SRL images of *C. elegans*, cholesterol-rich organelles are located and lipid unsaturation is mapped. These results collectively demonstrate the potential of a multi-window SRS imaging system in a wide range of applications.

## ASSOCIATED CONTENT

**Supporting Information**.
Additional information: Supplementary_FOPO.docx
This material is available free of charge via the Internet at http://pubs.acs.org.

## AUTHOR INFORMATION

Corresponding Author

* E-mail: jxcheng@bu.edu. Phone: +1 (617) 353-1276.

Author Contributions

All authors have given approval to the final version of the manuscript. ‡These authors contributed equally.

Funding Sources

NSF Chemical Imaging Program CHE1807106, NIH R35 GM136223, R01 AI141439, R01 CA224275 to J.X.C.

## ACKNOWLEDGMENT

We thank Professor Carlo Liberale of King Abdullah University of Science and Technology for helpful discussion on auto-balance detection. We acknowledge Dr. Maximilian Brinkmann and Dr. Tim Hellwig from Refined laser company for the assistance in FOPO usage and discussion on FOPO issues. We also acknowledge Haonan Lin's help in spectral unmixing algorithms. The ovarian cancers are provided by the Daniela Matei lab at Northwestern University.